\documentclass[runningheads]{llncs}
\usepackage{graphicx}

\usepackage{tikz}
\usepackage{comment}
\usepackage{amsmath,amssymb} 
\usepackage{color}

\usepackage[accsupp]{axessibility}  

\usepackage[width=122mm,left=12mm,paperwidth=146mm,height=193mm,top=12mm,paperheight=217mm]{geometry}

\begin{document}
\pagestyle{headings}
\mainmatter
\def\ECCVSubNumber{2}  

\title{PVT-COV19D: Pyramid Vision Transformer for COVID-19 Diagnosis} 

\titlerunning{ECCV-22 submission ID \ECCVSubNumber} 
\authorrunning{ECCV-22 submission ID \ECCVSubNumber} 
\author{Lilang Zheng\inst{1\#} \and Jiaxuan Fang\inst{1\#} \and Xiaorun Tang\inst{1} \and Hanzhang Li\inst{1} \and Jiaxin Fan\inst{1} \and Tianyi Wang\inst{1} \and Rui Zhou\inst{1*} \and Zhaoyan Yan\inst{1*}}
\institute{$ ^{1} $School of Information Science \& Engineering, Lanzhou University, China. \\
	\email{zr@lzu.edu.cn,yanchy16@lzu.edu.cn}\\
	($ ^{\#} $Equal contribution,*Co-corresponding authors)
}

\maketitle

\begin{abstract}
With the outbreak of COVID-19, a large number of relevant studies have emerged in recent years. We propose an automatic COVID-19 diagnosis framework based on lung CT scan images, the PVT-COV19D. In order to accommodate the different dimensions of the image input, we first classified the images using Transformer models, then sampled the images in the dataset according to normal distribution, and fed the sampling results into the modified PVTv2 model for training. A large number of experiments on the COV19-CT-DB dataset demonstrate the effectiveness of the proposed method.
\end{abstract}

\section{Introduction}

The Coronavirus Disease 2019 (COVID-19) is a highly contagious disease that has become a global epidemic, posing a serious threat to humanity worldwide. In order to prevent further spread of COVID-19 and promptly treat infected patients, early detection and isolation are important for a successful response to the COVID-19 pandemic.

Studies have found that COVID-19 can form a layer of lesions in the human lungs, which seriously affects the human respiratory system. The medical imaging characteristics of chest X-rays can quickly detect whether or not to be infected with COVID-19. Imaging features of the chest can be obtained using methods such as CT scans (computed tomography) and X-rays. Compared to X-rays, CT scans perform better, for example, X-rays give a two-dimensional view, while CT scans are made up of 3D views of the lungs, which help to check for disease and its location; also, machines used for CT scans Already in almost every country, the use of chest CT scans to detect COVID-19 has attracted the attention of many researchers.

However, rapid and accurate detection of COVID-19 using chest CT scans requires the guidance of specific experts, and timely and accurate detection of COVID-19 is a challenging task for healthcare workers. As more people need to be tested for COVID-19, medical staff are overwhelmed, reducing their focus on correctly diagnosing COVID-19 cases and confirming results. Therefore, it is necessary to distinguish normal cases from non-coronavirus infection and coronavirus cases in time to pay more attention to COVID-19 infection cases. Using deep learning-based algorithms to classify patients into novel coronaviruses and non-novel coronaviruses, medical staff can quickly exclude non-novel coronavirus cases in the first step and allocate more medical resources to more cases. Therefore, in order to reduce the human involvement of using chest CT images to detect COVID-19, an automated diagnostic model for detecting COVID-19 using imaging modalities is urgently needed.

In the past research, due to the breakthrough success of deep learning in the field of image recognition, many researchers have applied deep learning methods to the medical field \cite{kollias2018deep}. Combined with the latest research progress in big data analysis of medical images, Lei Cai et al. introduced the application of intelligent imaging and deep learning in the field of big data analysis and early diagnosis of diseases, especially the classification and segmentation of medical images \cite{cai2020review}. Amjad Rehman et al. introduced deep learning-based detection of COVID-19 using CT and X-ray images and data analysis of global spread \cite{9464121}. RAJIT NAIR et al. developed a fully automated model to predict COVID-19 using chest CT scans. The performance of the proposed method was evaluated by classifying CT scans of community-acquired pneumonia (CAP) and other non-pneumonic cases. The proposed deep learning model is based on ResNet 50, named CORNet, for detection of COVID-19, and a retrospective and multicenter analysis was also performed to extract visual features from volumetric chest CT scans during COVID-19 detection \cite{nair2021deep}.

For a long time, various computer vision tasks have been developed by using CNNs as the backbone. This paper proposes a simple CNN-free backbone for dense prediction tasks, the PVT-COV19D. PVT introduces a pyramid structure to the transformer, which enables various dense prediction tasks such as detection, segmentation, etc. In this paper, the method is applied to COVID-19 detection.

\section{Related Work}
After the outbreak of the COVID-19 pandemic, some studies have shown that deep learning method may be a potential method to detect COVID-19 \cite{2021Deep}, a lot of researches have used deep learning methods for diagnosis, mainly CNN on CT scan images or X-ray images. Unlike using CNN for classification, we use a simple backbone without CNN for intensive prediction tasks — Pyramid Vision Transformer(PVT) \cite{2021Pyramid}. 

\subsection{CNN}
In the past decade, compared with traditional machine learning and computer vision technology, deep learning has achieved the most advanced image recognition tasks. Similarly, the deep learning correlation scheme is widely used in the field of medical signal processing \cite{kollias2018deep}. We summarize the previous methods used for COVID-19 identification tasks. In \cite{C+R3} CNN+RNN network was used to input CT scanning images to distinguish COVID-19 and non-COVID cases. In \cite{3-D9}, the 3-D ResNet models was used to detect COVID-19, and the authors used volumetric 3-D CT scanning to distinguish it from other common pneumonia (CP) and normal cases. Wang et al. \cite{2020A} used 3D DeCoVNet to detect COVID-19, which took the CT volume of its lung mask as the input. In \cite{kollias2020transparent} D. Kollias et al. Classification of medical images using transparent adaptation (TA) method. In \cite{kollias2020deep} D. Kollias et al. proposed a method that extracts the potential information from the trained deep neural network (DNN), derives the concise representation, and analyzes it in an effective and unified way to achieve the purpose of prediction.

Among these classification methods, some use 2-D networks to predict slice images. This is called a 2-D network. In order to judge the lung CT, some methods use 2-D networks to generate embedded feature vectors for each image, then merge all feature vectors into global feature vectors and classify them using some fully connected (FC) layers. This is known as a 2-D plus 1-D network \cite{2+1D15,2+1D3}. Besides, voting methods are usually used \cite{voting2,voting17,voting8}. The third method is pure 3-D CNN network, which does not require slice annotation. It uses one or all available slice images as input, and the 3-D network processes all of these input images at once in the 3D channel space \cite{2020A} and \cite{3-D7}.

\subsection{The Vision Transformer}
The weight of convolution kernel is fixed after training, so it is difficult to adapt to the change of input. Therefore, some methods based on self-attention have been proposed to alleviate this problem. For example, the classic Non-local \cite{Non70}, which attempts to model the long-distance dependence in time domain and space domain, improves the accuracy of video classification; CCNet \cite{CCNet26} proposed cross-attention to reduce the amount of computation in Non-local; and stand-alone self-attention \cite{selfatten48} tried to replace convolution layer with local self-attention unit; AANet \cite{AANet3} combines self-attention and convolution; DERT \cite{DERT6} used transformer decoder to model target detection as an end-to-end dictionary query problem, and successfully removed post-processing such as NMS. Based on DERT, deformable DERT \cite{dDERT85} further introduces a deformable attention layer to focus on the sparse set of context elements, which makes convergence faster and performance better.

Different from the mature CNN, the backbone of the vision transformer is still in its early stage of development. Recently, ViT \cite{ViT13} established an image classification model using pure transformer, which takes a group of patches as image input; DeiT \cite{DeiT63} further extended ViT through a new distillation method. T2T ViT \cite{T2T37} gradually connect the marks in the overlapping sliding window into one mark. TNT \cite{TNT11} uses internal and external transformer blocks to generate pixel embedding and patch embedding  respectively. Because the characteristic graph of ViT output is single-scale, even the normal input size will consume a lot of computing overhead and memory overhead for ViT. Different from these methods, PVTv1 \cite{2021Pyramid} tried to introduce the pyramid structure into the transformer and designed a pure transformer backbone for intensive prediction tasks. It is the first transformer with pyramid structure. ViT and PVTv1 encode the image with 4*4 patch, and process the image into a set of non overlapping patch sequences, ignoring some local continuity of the image. In addition, both ViT and PVTv1 adopt fixed size position coding, which is not friendly to any size image. The computational complexity of PVTv1 is still very high. Therefore, Wang et al. proposed PVTv2 \cite{PVTv2*0} to improve these problems. It can get more local continuous images and feature maps, has the same linear complexity as CNN, and allows more flexible input.

\section{Methodology}

To adapt the different dimensional inputs of image, we can use a Transformer model for image classification by treating an image as a sequence of patches. The pyramid Transformer framework is to be used to generate multi-scale feature maps for the classification tasks. Our method use four stags to generate different scales of feature maps, which shares the similar architecture and it consists of a patch embedding layer and Transformer layer.

\begin{figure}
\centering
\includegraphics[height=6.5cm]{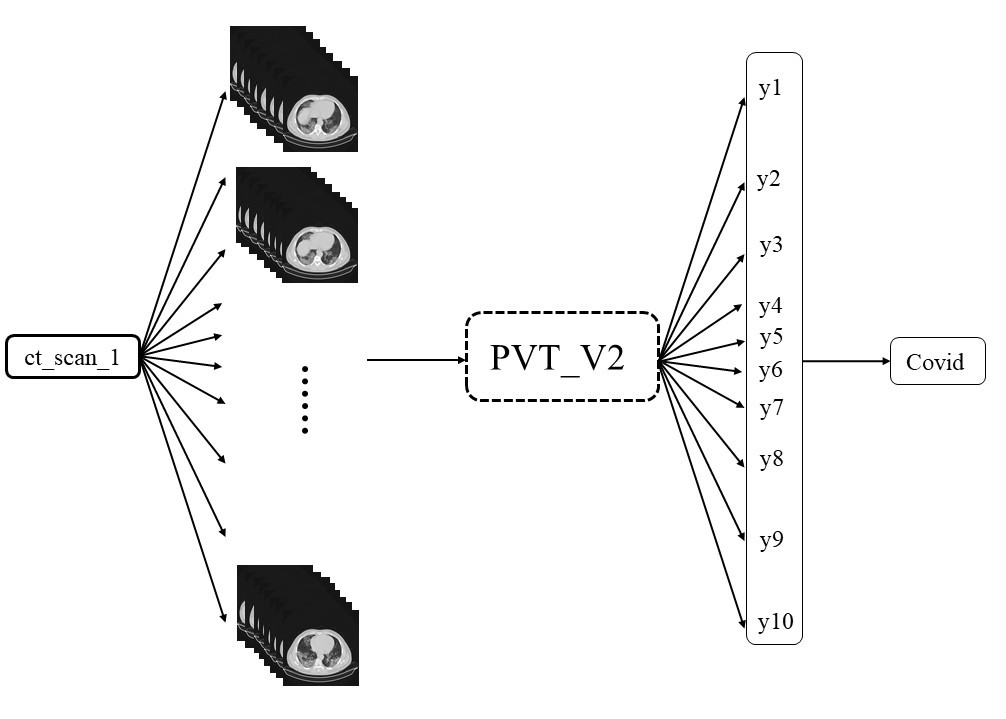}
\caption{Overall architecture of our model. Sampling from the dataset according to the normal distribution as iuput image. Use weighted average to judge the results}
\label{fig1}
\end{figure}

PVTv2 can obtain more local continuity of images and feature maps it can handle variable resolution inputs more flexibly while having the same linear complexity as CNN. Simply stack multiple independent Transformer encoders and gradually reduce the input resolution through Patch Embedding in each Stage. Our model samples the pictures in the dataset according to the normal distribution, 10 CT pictures of a person are sampled, each with 8 pictures. 

Each input 8 pictures into the model, these pictures can well reflect the patient's lung information, these pictures are input into the PVTv2 model, Multi-Head Attention can be roughly regarded as a global receptive field, and the result is According to the weighted average convolution of the attention weight, the Transformer's feature expression ability will be stronger.

\section{Dataset}

The COVID19-CT-Database (COV19-CT-DB) consists of chest CT scans that are annotated for the existence of COVID-19. Data were aggregated from many hospitals, containing anonymized human lung CT scans with signs of COVID-19 and without signs of COVID-19. The COV19-CT-DB database consist of about 7750 chest CT scan series. It consists of 1,650 COVID and 6,100 non-COVID chest CT scan series. Annotation of each  CT slice has been performed by 4 very experienced (each with over 20 years of experience) medical experts; two radiologists and two pulmonologists. Labels provided by the 4 experts showed a high degree of agreement (around 98\%).

One difference of COV19-CT-DB from other existing datasets is its annotation by medical experts (labels have not been created as a result of just positive RT-PCR testing).

Each of the scans includes different number of slices, ranging from 50 to 700. The database has been split in training, validation and testing sets. The training set contains, in total, 1993 cases. These include 882 COVID-19 cases and 1110 Non-COVID-19 cases. The validation set consists of 484 cases, 215 of them represent COVID-19 cases and 269 of them represent Non-COVID-19 cases. Both include different numbers of CT slices per CT scan, ranging from 50 to 700 \cite{2021MIA}.

\section{Experiments}
\subsection{Data pre-processing}
At first, CT images were extracted from DICOM files. Due to some slices of CT scan might be useless for recognizing the COVID-19 (e.g., top/bottom slices might not contain chest information), the slices selection of the CT scan in the training phase is essential, as well as in the evaluation phase. During the training phase, CT scan sections were sampled according to normal distribution, a small number of CT scan sections were collected from the top and bottom, and more CT scan sections were collected from the central part with a higher possibility of lesion.And then these selected sections were enhanced and normalized. In image analysis, the quality of the image directly affects the design of the recognition algorithm and the accuracy of the effect, so before image analysis (feature extraction, segmentation, matching and recognition, etc.), it needs to be preprocessed. The main purpose of image preprocessing is to eliminate irrelevant information in images, recover useful real information, enhance the detectability of relevant information, simplify data to the maximum extent, and improve the reliability of classification. First, image enhancement is performed on the data \cite{son2020urie}. The main purpose of image enhancement is to improve the image quality and recognizability, so that the image is more conducive to observation or further analysis and processing. Image enhancement technology usually highlights or enhances some features of the image, such as edge information, contour information and contrast, so as to better display the useful information of the image and improve the use value of the image. Image enhancement technology is under a certain standard, the processed image is better than the original image. Then, normalized to the range of [0, 1].

\subsection{Implementation Details}
A lung CT scan may contain dozens to hundreds of pictures, Most of the useful information is concentrated in the middle of these pictures.Thus, we use normal sampling each time, randomly select a batch = 8 pictures in the CT scan of the same case as the input of the neural network, those pictures in the middle of a case are more likely to be input to the model for training,which makes the model to converge faster. 

We set the target of each positive CT scan to 1, and set the target of each negative CT scan to -1, use MSEloss as the loss function, and AdamW as the optimizer with an initial learning rate of $ 1\times 10^{-4}  $. In the evaluation stage, we introduced a multiple voting mechanism, inputting a batch into the model will get several outputs, and then the outputs will be averaged, and the plus or minus sign of this average will be used as one evaluation standard.the above steps will be performed n times. if more than half of the averages are plus, the case is considered as positive, otherwise it is negative.

\begin{table}[]
	\caption{Comparison of the proposed method with the baseline}
	\label{tab1}
	\begin{center}
		\begin{tabular}{cc}
			\hline
			Method        & Macro F1 Score \\ \hline
			Baseline      & 77.00\%          \\ \hline
			PVT-COV19D(ours) & 88.01\%          \\ \hline   
		\end{tabular}
	\end{center}
\end{table}

Our models are trained with a batch size of 8 on one 2080Ti GPU, the training and validation set are partitioned by~\cite{2021MIA}, where the number of training and validation CT scans are 1, 992 and 484. After training for 60 epochs, the positive accuracy of our model reached 84.11\%, the negtive accuracy reached 91.54\%, and the validation set prediction accuracy reached 88.19\%, the macro F1 score reached 0.8801 for this binary classification, The performance comparison between the proposed and baseline~\cite{kollias2022ai} is showed in Table~\ref{tab1}.

\section{Conclusions}

In this paper we present a PVT-COV19D network for COVID-19 classification. This network is different from CNN, it can get more local continuous images and feature maps, has the same linear complexity as CNN, and allows more flexible input. In addition, in the data processing stage, we added random sampling to speed up the convergence of the model. Experiments show that our introduced random sampling strategy and PVTv2 framework achieve a good balance between speed and accuracy and our network requires less computing power for COVID-19 detection.



%
%
\bibliographystyle{splncs04}
\bibliography{egbib}

\begin{thebibliography}{10}
\providecommand{\url}[1]{\texttt{#1}}
\providecommand{\urlprefix}{URL }
\providecommand{\doi}[1]{https://doi.org/#1}

\bibitem{AANet3}
Bello, I., Zoph, B., Le, Q., Vaswani, A., Shlens, J.: Attention augmented
  convolutional networks. In: 2019 IEEE/CVF International Conference on
  Computer Vision (ICCV) (2020)

\bibitem{cai2020review}
Cai, L., Gao, J., Zhao, D.: A review of the application of deep learning in
  medical image classification and segmentation. Annals of translational
  medicine  \textbf{8}(11) (2020)

\bibitem{DERT6}
Carion, N., Massa, F., Synnaeve, G., Usunier, N., Kirillov, A., Zagoruyko, S.:
  End-to-end object detection with transformers. In: European conference on
  computer vision. pp. 213--229. Springer (2020)

\bibitem{voting2}
Chaudhary, S., Sadbhawna, Jakhetiya, V., Subudhi, B.N., Guntuku, S.C.:
  Detecting covid-19 and community acquired pneumonia using chest ct scan
  images with deep learning. In: ICASSP 2021 - 2021 IEEE International
  Conference on Acoustics, Speech and Signal Processing (ICASSP) (2021)

\bibitem{ViT13}
Dosovitskiy, A., Beyer, L., Kolesnikov, A., Weissenborn, D., Zhai, X.,
  Unterthiner, T., Dehghani, M., Minderer, M., Heigold, G., Gelly, S., et~al.:
  An image is worth 16x16 words: Transformers for image recognition at scale.
  arXiv preprint arXiv:2010.11929  (2020)

\bibitem{2+1D3}
Garg, P., Ranjan, R., Upadhyay, K., Agrawal, M., Deepak, D.: Multi-scale
  residual network for covid-19 diagnosis using ct-scans. In: ICASSP 2021 -
  2021 IEEE International Conference on Acoustics, Speech and Signal Processing
  (ICASSP) (2021)

\bibitem{TNT11}
Han, K., Xiao, A., Wu, E., Guo, J., Xu, C., Wang, Y.: Transformer in
  transformer. Advances in Neural Information Processing Systems  \textbf{34},
  15908--15919 (2021)

\bibitem{3-D7}
He, X., Wang, S., Chu, X., Shi, S., Tang, J., Liu, X., Yan, C., Zhang, J.,
  Ding, G.: Automated model design and benchmarking of 3d deep learning models
  for covid-19 detection with chest ct scans. arXiv preprint arXiv:2101.05442
  (2021)

\bibitem{voting8}
Heidarian, S., Afshar, P., Enshaei, N., Naderkhani, F., Rafiee, M.J.,
  Babaki~Fard, F., Samimi, K., Atashzar, S.F., Oikonomou, A., Plataniotis,
  K.N., et~al.: Covid-fact: A fully-automated capsule network-based framework
  for identification of covid-19 cases from chest ct scans. Frontiers in
  Artificial Intelligence  \textbf{4},  598932 (2021)

\bibitem{CCNet26}
Huang, Z., Wang, X., Wei, Y., Huang, L., Huang, T.S.: Ccnet: Criss-cross
  attention for semantic segmentation. IEEE Transactions on Pattern Analysis
  and Machine Intelligence  \textbf{PP}(99), ~1--1 (2020)

\bibitem{kollias2022ai}
Kollias, D., Arsenos, A., Kollias, S.: Ai-mia: Covid-19 detection \& severity
  analysis through medical imaging. arXiv preprint arXiv:2206.04732  (2022)

\bibitem{2021MIA}
Kollias, D., Arsenos, A., Soukissian, L., Kollias, S.: Mia-cov19d: Covid-19
  detection through 3-d chest ct image analysis. In: Proceedings of the
  IEEE/CVF International Conference on Computer Vision. pp. 537--544 (2021)

\bibitem{kollias2020deep}
Kollias, D., Bouas, N., Vlaxos, Y., Brillakis, V., Seferis, M., Kollia, I.,
  Sukissian, L., Wingate, J., Kollias, S.: Deep transparent prediction through
  latent representation analysis. arXiv preprint arXiv:2009.07044  (2020)

\bibitem{kollias2018deep}
Kollias, D., Tagaris, A., Stafylopatis, A., Kollias, S., Tagaris, G.: Deep
  neural architectures for prediction in healthcare. Complex \& Intelligent
  Systems  \textbf{4}(2),  119--131 (2018)

\bibitem{kollias2020transparent}
Kollias, D., Vlaxos, Y., Seferis, M., Kollia, I., Sukissian, L., Wingate, J.,
  Kollias, S.D.: Transparent adaptation in deep medical image diagnosis. In:
  TAILOR. p. 251–267 (2020)

\bibitem{2+1D15}
Li, L., Qin, L., Xu, Z., Yin, Y., Wang, X., Kong, B., Bai, J., Lu, Y., Fang,
  Z., Song, Q., et~al.: Artificial intelligence distinguishes covid-19 from
  community acquired pneumonia on chest ct. Radiology  (2020)

\bibitem{3-D9}
Li, Y., Xia, L.: Coronavirus disease 2019 (covid-19): Role of chest ct in
  diagnosis and management. American Journal of Roentgenology  \textbf{214}(6),
  ~1--7 (2020)

\bibitem{voting17}
Mr, A., Aa, B., Sms, C.: A fully automated deep learning-based network for
  detecting covid-19 from a new and large lung ct scan dataset - sciencedirect.
  Biomedical Signal Processing and Control  \textbf{68}

\bibitem{nair2021deep}
Nair, R., Alhudhaif, A., Koundal, D., Doewes, R.I., Sharma, P.: Deep
  learning-based covid-19 detection system using pulmonary ct scans. Turkish
  Journal of Electrical Engineering and Computer Sciences  \textbf{29}(8),
  2716--2727 (2021)

\bibitem{C+R3}
Natarajan, Y., Khadidos, A., Tsaramirsis, G., Khadidos, A.O., Mohanty, S.N.:
  Analysis of covid-19 infections on a ct image using deepsense model.
  Frontiers in Public Health  \textbf{8}, ~1--9 (2020)

\bibitem{selfatten48}
Ramachandran, P., Parmar, N., Vaswani, A., Bello, I., Levskaya, A., Shlens, J.:
  Stand-alone self-attention in vision models. In: Neural Information
  Processing Systems (2019)

\bibitem{9464121}
Rehman, A., Saba, T., Tariq, U., Ayesha, N.: Deep learning-based covid-19
  detection using ct and x-ray images: Current analytics and comparisons. IT
  Professional  \textbf{23}(3),  63--68 (2021). \doi{10.1109/MITP.2020.3036820}

\bibitem{son2020urie}
Son, T., Kang, J., Kim, N., Cho, S., Kwak, S.: Urie: Universal image
  enhancement for visual recognition in the wild. In: ECCV (2020)

\bibitem{2021Deep}
Song, Y., Zheng, S., Li, L., Zhang, X., Yang, Y.: Deep learning enables
  accurate diagnosis of novel coronavirus (covid-19) with ct images. IEEE/ACM
  Transactions on Computational Biology and Bioinformatics  \textbf{PP}(99),
  ~1--1 (2021)

\bibitem{DeiT63}
Touvron, H., Cord, M., Douze, M., Massa, F., Sablayrolles, A., J{\'e}gou, H.:
  Training data-efficient image transformers \& distillation through attention.
  In: International Conference on Machine Learning. pp. 10347--10357. PMLR
  (2021)

\bibitem{2021Pyramid}
Wang, W., Xie, E., Li, X., Fan, D.P., Song, K., Liang, D., Lu, T., Luo, P.,
  Shao, L.: Pyramid vision transformer: A versatile backbone for dense
  prediction without convolutions. In: Proceedings of the IEEE/CVF
  International Conference on Computer Vision. pp. 568--578 (2021)

\bibitem{PVTv2*0}
Wang, W., Xie, E., Li, X., Fan, D.P., Song, K., Liang, D., Lu, T., Luo, P.,
  Shao, L.: Pvt v2: Improved baselines with pyramid vision transformer.
  Computational Visual Media  \textbf{8}(3),  415--424 (2022)

\bibitem{2020A}
Wang, X., Deng, X., Fu, Q., Zhou, Q., Zheng, C.: A weakly-supervised framework
  for covid-19 classification and lesion localization from chest ct. IEEE
  Transactions on Medical Imaging  \textbf{PP}(99), ~1--1 (2020)

\bibitem{Non70}
Wang, X., Girshick, R., Gupta, A., He, K.: Non-local neural networks. In:
  Proceedings of the IEEE conference on computer vision and pattern
  recognition. pp. 7794--7803 (2018)

\bibitem{T2T37}
Yuan, L., Chen, Y., Wang, T., Yu, W., Shi, Y., Jiang, Z.H., Tay, F.E., Feng,
  J., Yan, S.: Tokens-to-token vit: Training vision transformers from scratch
  on imagenet. In: Proceedings of the IEEE/CVF International Conference on
  Computer Vision. pp. 558--567 (2021)

\bibitem{dDERT85}
Zhu, X., Su, W., Lu, L., Li, B., Wang, X., Dai, J.: Deformable detr: Deformable
  transformers for end-to-end object detection. arXiv preprint arXiv:2010.04159
   (2020)

\end{thebibliography}
\end{document}